\documentclass{aa}
\usepackage{txfonts}
\usepackage{graphicx}
\begin{document}
\title{Flux-limited strong gravitational lensing and dark energy}
\author{Da-Ming Chen}
\institute{National Astronomical Observatories, Chinese Academy of
Sciences, Beijing 100012, China} \offprints{Da-Ming Chen,
\email{cdm@bao.ac.cn}}
\date{Received 23 June 2003 / Accepted 15 January 2004}

\abstract{In the standard flat cosmological constant ($\Lambda$)
cold dark matter (CDM) cosmology, a model of two populations of
lens halos for strong gravitational lensing can reproduce the
results of the Jodrell-Bank VLA Astrometric Survey (JVAS) and the
Cosmic Lens All-Sky Survey (CLASS) radio survey. In such a model,
lensing probabilities are sensitive to three parameters: the
concentration parameter $c_1$, the cooling mass scale
$M_\mathrm{c}$ and the value of the CDM power spectrum
normalization parameter $\sigma_8$. The value ranges of these
parameters are constrained by various observations. However, we
found that predicted lensing probabilities are also quite
sensitive to the flux density (brightness) ratio $q_{\mathrm{r}}$
of the multiple lensing images, which has been a very important
selection criterion of a sample in any lensing survey experiments.
We re-examine the mentioned above model by considering the flux
ratio and galactic central Super Massive Black Holes (SMBHs), in
flat, low-density cosmological models with different cosmic
equations of state $\omega$, and find that the predicted lensing
probabilities without considering $q_{\mathrm{r}}$ are
over-estimated.  A low value of $q_\mathrm{r}$ can be compensated
for by raising the cooling mass scale $M_\mathrm{c}$ in fitting
the predicted lensing probabilities to JVAS/CLASS observations. To
determine the cosmic equation of state $\omega$, the uncertainty
in $M_\mathrm{c}$ must be resolved. The effects of SMBHs cannot be
detected by strong gravitational lensing methods when
$q_{\mathrm{r}}\leq 10$. }
\titlerunning{Flux-limited strong gravitational lensing and dark energy}
\authorrunning{D.-M. Chen}
\maketitle
\section{Introduction}
\label{intro} Gravitational lensing provides us with a powerful
probe of mass distribution of the universe. By comparing the
lensing probabilities predicted by various cosmological models and
the density profile of lenses with observations, we are able to
test the mass distribution of dark matter halos and, in
particular, the inner density slope because the Jodrell-Bank VLA
Astrometric Survey (JVAS) and the Cosmic Lens All-Sky Survey
(CLASS; Browne et al.~\cite{browne}; Helbig~\cite{helbig}; Browne
et al.~\cite{browne02}; Myers et al.~\cite{myers}) has provided us
with observed lensing probabilities at small image separations
($0.3''<\Delta\theta<3''$).

In a standard flat $\Lambda$CDM cosmology, it is believed that a
model of two populations of lens halos for strong gravitational
lensing can reproduce the results of the JVAS/CLASS radio survey
(Porciani \& Madau~\cite{porci}; Kochanek \& White~\cite{kocha};
Keeton~\cite{keeton2001a}; Keeton \& Madau~\cite{keeton2001b};
Sarbu, Rusin \& Ma~\cite{sarbu}; Li \& Ostriker~\cite{li};
Oguri~\cite{oguri}; Oguri~\cite{oguri2003a}; Oguri, Lee \&
Suto~\cite{oguri2003b}). The two populations of lens halos are
distinguished by introducing an abrupt change in the structure of
the objects at the cooling mass scale $M_{\mathrm{c}}$ dividing
galaxies and clusters. Some authors used Singular Isothermal
Sphere (SIS) models below $M_{\mathrm{c}}$ and Navarro-Frenk-White
(NFW) models above $M_{\mathrm{c}}$ and could fit the observed
image separation distribution of JVAS/CLASS (Li \&
Ostriker~\cite{li}; Sarbu, Rusin \& Ma~\cite{sarbu}). Based on the
fact that the image separation distribution of lenses below
$1^{''}$ depends sensitively on both the inner mass profile of
galactic halos and the faint end slope of the mass and luminosity
functions, Ma (\cite{ma2003}) compared the traditional approach
that models the lenses as SIS and the Schechter luminosity
function with a dark matter based approach that models the lenses
with a certain halo mass profile and the Press-Chechter mass
function. Constraints on the inner total mass profiles of halos
are investigated by requiring the two approaches to give
consistent predictions.  Li \& Ostriker~(\cite{li2003}) further
proposed a model of three populations of halos as lenses
distinguished by the halos mass to calculate the lensing
probability of image separation and time delay.

In this paper, we revisit the model of the two populations of lens
halos to calculate the lensing probabilities in a flat
quintessence cold dark matter (QCDM) cosmology with different
cosmic equations of state $\omega$. The focus here is on the flux
density ratio $q_{\mathrm{r}}$ and cooling mass scale
$M_{\mathrm{c}}$, and their effects on the estimate of the
equation of state of dark energy $\omega$. In spite of the fact
that low mass lensing halos (galactic size or less) cannot all be
SIS (Ma~\cite{ma2003}; Li \& Ostriker~\cite{li2003}; Benson et
al.~\cite{benson}; Somerville \& Primack~\cite{somer}; Kaufmann et
al.~\cite{kaufm}), we still model the galactic lens halos as SIS
in our calculations based on the following considerations. On the
one hand, evidence based on stellar dynamics of elliptical
galaxies (e.g., Rix et al.~\cite{rix}; Romanowsky \&
Kochanek~\cite{roman}; Treu \& Koopmans~\cite{treu}), modeling of
lensed systems (e.g., Cohn et al.~\cite{cohn}) and flux ratios of
multiple images (Rusin \& Ma~\cite{rusin2001};
Rusin~\cite{rusin2002}) all give an inner profile for lensing
galaxies that is consistent with SIS. On the other hand, since the
SIS lensing cross section is several orders of magnitude higher
than that of the NFW, the lensing probabilities depend strongly on
the cooling mass scale $M_{\mathrm{c}}$ because a larger
$M_{\mathrm{c}}$ (and a smaller $M_{\mathrm{c}}$ at the lower mass
end of SIS lens halos, see Li \& Ostriker~\cite{li2003}) allows
more halos to be modeled as SIS. So the importance of the
subdivision for low mass lens halos is lessened unless the large
uncertainty of $M_{\mathrm{c}}$ is sufficiently reduced. We thus
ignore such a subdivision in this paper, and focus on the roles
played by the flux density ratio $q_{\mathrm{r}}$ and cooling mass
scale $M_{\mathrm{c}}$ in constraining the equation of state
$\omega$.

In addition to $q_{\mathrm{r}}$ and $M_{\mathrm{c}}$, lensing
probabilities estimated by the NFW halo are also sensitive to the
concentration parameter $c_1$. We use the mass-dependent fitting
formula given by Bullock et al.~(\cite{bullo}). For a given halo
mass and redshift, there is a scatter in $c_1$ that is consistent
with a log-normal distribution with standard deviation
$\sigma_c=\Delta(\log c)\approx 0.18$~(Jing\cite{jing}; Bullock et
al.~\cite{bullo}). Taking into account the scatter in $c_1$ by
averaging the lensing probability with the lognormal distribution
will increase the probabilities considerably at larger image
separations and only slightly at smaller separations
(Chen~\cite{chenb}). Since the well-defined sample of JVAS/CLASS
is limited to small image separations ($0.3''<\Delta\theta<3''$),
we thus ignore the scatter in $c_1$ and our conclusions will not
be affected. Another major uncertainty in predicting lensing
probability arises from the considerable uncertainty regarding the
value of the CDM power spectrum normalization parameter $\sigma_8$
(Chen~\cite{chenb}); we adopt $\sigma_8=0.95$, suggested by the
abundance of massive clusters at redshifts $z=0.5\sim 0.8$
(Bahcall \& Bode~\cite{bahcall}) which gives $\sigma_8=0.98\pm
0.1$. Lensing probability increases quickly with the source
redshift $z_s$; since its distribution in the JVAS/CLASS survey is
still poorly understood, we use the estimated mean value of
$<z_s>=1.27$ (Marlow et al.~\cite{marlow}).

On the other hand, it should be pointed out that, like most
authors, we use the spherical lensing profiles. Non-spherical
lensing profiles do not change significantly the separation of
observed multiple images, unless the profile deviates
significantly from the spherical one. However, as pointed out by
many authors (e.g. in a cluster environment, this has been
discussed by Meneghetti et al.~\cite{meneg2003a,meneg2003b}; for
the importance of substructures see Mao \& Schneider~\cite{mao}),
the elliptical lenses and their substructures do have
significantly larger cross sections. A more realistic lensing
model should include these effects, which will be discussed in
another paper.

So in this paper we use the model of two populations of lens halos
mentioned above to investigate the effect of image flux density
ratio $q_{\mathrm{r}}$ on lensing probabilities and find that the
predicted lensing probabilities without considering
$q_{\mathrm{r}}$ are over-estimated. In flat, low-density
cosmological models with different cosmic equations of state
$\omega$, we show that when the flux density ratio $q_\mathrm{r}$
changes from $10$ to $10^4$, the corresponding fit value of the
cosmic equation of state  will change from $\omega=-1$ to
$\omega=-1/2$. Also, in our flux-limited statistics of strong
lensing image separations, the contributions from galactic central
SMBHs can be safely ignored.

The paper is organized as follows: Sect.~\ref{cosmology} provides
a brief description of the cosmology model used in this paper,
Sect.~\ref{lensprob} presents the predicted integral lensing
probabilities, Sect.~\ref{discuss} gives our results, the
discussion and comparisons with previous work. Appendix A is
devoted to a detailed deduction of lensing equations for galactic
halos.

\section{Cosmological model and mass function}
\label{cosmology}

In this section, we describe the cosmological model and dark halo
mass function.  More and more evidence shows that the Universe at
present is dominated by a smooth component with negative pressure,
the so called dark energy. So far, two kinds of dark energy have
been proposed, quintessence and the so-called Chaplygin gas (e.g.,
Bento, Bertolami \& Sen~\cite{bento}, \cite{bento03}). We consider
only the quintessence in this paper. It is assumed that
quintessence offers an alternative to the cosmological constant as
the missing energy in a spatially flat universe with a subcritical
matter density $\Omega_\mathrm{m}$ (Caldwell, Dave \&
Steinhardt~\cite{caldw}). In this paper, we study spatially flat
QCDM models in which the cold dark matter and quintessence-field
make up the critical density (i.e.,
$\Omega_\mathrm{m}+\Omega_\mathrm{Q}=1$). Throughout the paper, we
choose the most generally accepted values of the parameters for
flat, QCDM cosmology, in which, with the usual symbols, the matter
density parameter, dark energy density parameter and Hubble
constant are respectively: $\Omega_{\mathrm m}=0.3$,
$\Omega_{\mathrm{Q}}=0.7$, $h=0.75$. Four negative values of
$\omega$ in equation of state
$p_\mathrm{Q}=\omega\rho_\mathrm{Q}$, with $\omega=-1$
(cosmological constant), $\omega=-2/3$, $\omega=-1/2$ and
$\omega=-1/3$, are chosen to see their effects on lensing
probabilities. We use the conventional form to express the
redshift $z$-dependent linear power spectrum for the matter
density perturbation in a QCDM cosmology established by Ma et al.
(\cite{ma1999})
\begin{equation}
P(k,z)=A_\mathrm{Q}k^nT^2_\mathrm{Q}(k,z)
\left(\frac{g_\mathrm{Q}(z)}{g_\mathrm{Q}(0)(1+z)}\right)^2.
\label{spectrum}
\end{equation}
The normalization $A_\mathrm{Q}$ can be determined by $\sigma_8$
for different values of $\omega$
\begin{equation}
\sigma_8^2=\frac{1}{2\pi^2}\int^{\infty}_0P(k,0) W^2(kr_8)k^2dk,
\end{equation}
in which $W(x)=3[\sin(x)/x^3-\cos(x)/x^2]$ is the Fourier
transformation of a top-hat window function, and
$r_8=8h^{-1}\mathrm{Mpc}$. We choose the spectral index of the
primordial adiabatic density perturbations to be $n=1$.
$T_\mathrm{Q}$ is the transfer function for QCDM models, which is
related to $T_{\Lambda}$, the transfer function for $\Lambda$CDM
models, with the relative transfer function
\begin{equation}
T_{\mathrm{Q}\Lambda}=\frac{T_\mathrm{Q}}{T_{\Lambda}}=\frac{\alpha+\alpha
q^2}{1+\alpha q^2}, \ \ \ \ q=\frac{k}{\Gamma_\mathrm{Q}h},
\end{equation}
where $\Gamma_\mathrm{Q}$ is the shape parameter in QCDM models
and $\alpha$ is a coefficient that quantifies the relative
amplitude of the matter density field on large and small length
scales, and is well approximated by
\begin{eqnarray}
\alpha&=&(-\omega)^s,  \nonumber \\
s&=&(0.012-0.036\omega-0.017/\omega)[1-\Omega_\mathrm{m}(z)]
\nonumber \\
& &+(0.098+0.029\omega-0.085/\omega) \ln\Omega_\mathrm{m}(z),
\end{eqnarray}
where $\Omega_\mathrm{m}(z)=\Omega_\mathrm{m}/[\Omega_\mathrm{m}
+(1-\Omega_\mathrm{m})(1+z)^{3\omega}]$ is the matter density
parameter. We use the fitting formula for $\Lambda$CDM transfer
function given by Eisenstein \& Hu (\cite{eisen}):
$T_{\Lambda}=L/(L+Cq_\mathrm{eff}^2)$, with
$L=\ln(e+1.84q_\mathrm{eff})$,
$q_\mathrm{eff}=k/(\Omega_\mathrm{m}h^2\mathrm{Mpc}^{-1})$ and
$C=14.4+325/(1+60.5q_\mathrm{eff}^{1.11})$.

Similarly, the linear growth suppression factor of the density
field $g_\mathrm{Q}$ for QCDM in equation (\ref{spectrum}) is
related to that for $\Lambda$CDM $g_{\Lambda}$ with
\begin{eqnarray}
g_\mathrm{Q\Lambda}&=&\frac{g_\mathrm{Q}}{g_{\Lambda}}=(-\omega)^{-t},
\nonumber \\
t&=&-(0.255+0.305\omega+0.0027/\omega)[1-\Omega_\mathrm{m}(z)]
\nonumber \\
& &-(0.366+0.266\omega-0.07/\omega)\ln\Omega_\mathrm{m}(z),
\end{eqnarray}
where the empirical fit of $g_{\Lambda}$ is given by
$g_{\Lambda}=2.5\Omega_\mathrm{m}(z)/\{\Omega_\mathrm{m}(z)^{4/7}
-1+\Omega_\mathrm{m}(z)+[1+\Omega_\mathrm{m}(z)/2]
[1+(1-\Omega_\mathrm{m}(z))/70]\}$ (Carroll, Press \&
Turner~\cite{carro}).

We know that most of the consequences of quintessence follow from
its effect on evolution of the expansion rate of the Universe,
which in turn affects the growth of density perturbations (as
described above) and the cosmological distances. From the first
Friedmann equation $d(\rho_\mathrm{Q}a^3)=-p_\mathrm{Q}da^3$
($a=1/(1+z)$ is the scale factor of the Universe), the expansion
rate (Hubble constant) for a flat Universe and then the proper
distance and the angular-diameter distance can be calculated in
our QCDM models (Huterer \& turner~\cite{huter}), which are needed
in predicting lensing probabilities.

The physical number density $\bar{n}(M,z)$ of virialized dark
halos of masses between $M$ and $M+dM$ is related to the comoving
number density $n(M,z)$ by $\bar{n}(M,z)=n(M,z)(1+z)^3$, the
latter originally given by Press \& Schechter (\cite{press74}),
and the improved version is $n(M,z)dM=(\bar{\rho}(0)/M)f(M,z)dM$,
where $\bar{\rho}(0)$ is the current mean mass density of the
universe, and $f(M,z)$ is the mass function for which we use the
expression given by Jenkins et al. (\cite{jenki}).

\section{Lensing probabilities}
\label{lensprob}

In this section, we first give a brief description of lensing
equations for galaxies and clusters of galaxies, then present the
predicted integral lensing probabilities. According to the model
of two populations of halos, cluster-size halos are modeled as NFW
profile: $\rho_\mathrm{NFW}=\rho_\mathrm{s}r_\mathrm{s}^3/
[r(r+r_\mathrm{s})^2]$ , where $\rho_\mathrm{s}$ and
$r_\mathrm{s}$ are constants. We can define the mass of a halo to
be the mass within the virial radius of the halo $r_\mathrm{ vir}$
: $M_\mathrm{DM}=4\pi\rho_\mathrm{s}r_\mathrm{s}^3f(c_1)$, where
$f(c_1)=\ln(1+c_1)-c_1/(1+c_1)$, and $c_1=r_\mathrm{
vir}/r_\mathrm{s}=9(1+z)^{-1}(M/1.5\times
10^{13}h^{-1}M_{\sun})^{-0.13}$ is the concentration parameter,
for which we have used the fitting formula given by Bullock et al.
(\cite{bullo}). The lensing equation for NFW lenses is as usual
$y=x-\mu_s g(x)/x$ (Li \& Ostriker~\cite{li};
Chen~\cite{chena,chenb}), where $y=|\vec{y}|$,
$\vec{\eta}=\vec{y}D_\mathrm{S}^\mathrm{A}/D^\mathrm{A}_\mathrm{L}$
is the position vector in the source plane, in which
$D_\mathrm{S}^\mathrm{A}$ and $D_\mathrm{L}^\mathrm{A}$ are
angular-diameter distances from the observer to the source and to
the lens respectively.  $x=|\vec{x}|$ and
$\vec{x}=\vec{\xi}/r_\mathrm{s}$, $\vec{\xi}$ is the position
vector in the lens plane. Since the surface mass density is
circularly symmetric, we can extend both $x$ and $y$ to their
opposite values in our actual calculations for convenience. The
parameter
$\mu_\mathrm{s}=4\rho_\mathrm{s}r_\mathrm{s}/\Sigma_\mathrm{cr}$
is $x$ independent, in which $\Sigma_\mathrm{cr}=(c^2/4\pi
G)(D_\mathrm{S}^\mathrm{A}/D_\mathrm{L}^\mathrm{A}
D_\mathrm{LS}^\mathrm{A})$ is the critical surface mass density,
with $c$ the speed of light, $G$ the gravitational constant and
$D_\mathrm{LS}^\mathrm{A}$ the angular-diameter distance from the
lens to the source. The function $g(x)$ has an analytical
expression originally given by Bartelmann (\cite{barte}).

Since the observational evidence presented so far suggests the
ubiquity of black holes in the nuclei of all bright galaxies
regardless of their activity (Magorrian et al. \cite{magor};
Ferrarese \& Merritt \cite{ferra}; Ravindranath et al.
\cite{ravin}; Merritt \& Ferrarese \cite{merria}, \cite{merrib},
\cite{merric}; Wandel \cite{wandel}; Sarzi et al. \cite{sarzi}),
we add a central supermassive black hole (SMBH) as a point mass to
each SIS modeled galactic lens halo in its
center(Keeton~\cite{keeton02}; Chen~\cite{chena}, \cite{chenb}).
It is well known that the finite  flux density ratio will
definitely reduce the value of the lensing cross
section(Schneider, Ehlers \& Falco~\cite{schne}). In the model of
one population of halos (NFW) combined with each galactic halo a
central point mass, the lensing probabilities are shown to be
sensitive to  the flux density ratio (Chen~\cite{chena},
\cite{chenb}). One may argue that this is the case only because we
treat the galactic bulge as a point mass (some fraction of the
bulge mass), since a point mass lens will always produce a larger
flux density ratio $q_{\mathrm{r}}$ of the multiple images. So it
would be interesting to investigate the model of two populations
of halos to see whether or not the lensing probabilities will be
sensitive to $q_{\mathrm{r}}$. We will show later that the answer
is yes, and the contribution from an individual central galactic
SMBH is detectable only if $q_{\mathrm{r}}\geq 10^3$. When
$q_{\mathrm{r}}\leq 10$ (for the well-defined sample of
JVAS/CLASS), the effect of SMBH can be safely ignored.

So, for galaxy-size lenses, we use an SIS+SMBH model. The surface
mass density is
\begin{equation}
\Sigma(\vec{\xi})=M_{\bullet}\delta^2(\vec{\xi})
+\frac{\sigma^2_{v}}{2G|\vec{\xi}|}, \label{xi0}
\end{equation}
where $M_{\bullet}$ is the mass of a SMBH, $\vec{\xi}$ is the
position vector in the lens plane, $\delta^2(\vec{\xi})$ is the
two dimensional Dirac-delta function, and $\sigma_{v}$ is the
velocity dispersion. Choosing certain length scales, the lensing
equation can be written as $y=x-x_\mathrm{h}/x-|x|/x$, where
$x_\mathrm{h}$ is the scaled radius of the sphere of influence of
a SMBH (see Appendix A for details).

When the quasars at the mean redshift $<z_{\mathrm{s}}>=1.27$ are
lensed by foreground CDM halos of galaxies and clusters of
galaxies, the lensing probability with image separations larger
than $\Delta\theta$ and flux density ratio less than
$q_{\mathrm{r}}$ is (Schneider et al. \cite{schne}) {\small
\begin{equation}
P(>\Delta\theta, <q_{\mathrm{r}})=
\int^{z_{\mathrm{s}}}_0\frac{dD_{\mathrm{L}}(z)}
{dz}dz\int^{\infty}_0\bar{n}(M,z)\sigma(M,z)B(M,z)dM,
\label{prob1}
\end{equation}
} where $D_{\mathrm{L}}(z)$ is the proper distance from the
observer to the lens located at redshift $z$. $\bar{n}(M,z)$ is
the physical number density of virialized dark halos of masses
between $M$ and $M+dM$ at redshift $z$ given by Jenkins et al.
(\cite{jenki}). The cross section $\sigma(M,z)$ is mass and
redshift dependent, and is sensitive to the flux density ratio of
multiple images $q_\mathrm{r}$ for SIS halos, as shown below in
detail. For the magnification bias $B(M,z)$, we use the results
given by Li \& Ostriker (\cite{li}) for both SIS+SMBH and NFW
lense. For the former, if the contribution from a SMBH is ignored,
the magnification bias for SIS only will be recovered (which is
$B_\mathrm{SIS}\approx 4.76$).

The cross section for the cluster-size NFW lenses has been well
studied (Li \& Ostriker~\cite{li}). The lensing equation is
$y=x-\mu_s g(x)/x$ and the multiple images can be produced only if
$|y|\leq y_{\mathrm{cr}}$, where $y_{\mathrm{cr}}$ is the maximum
value of $y$ when $x<0$, which is determined by $dy/dx=0$ and the
cross section in the lens plane is simply $\sigma(M, z)=\pi
y_\mathrm{cr}^2r_\mathrm{s}^2$. We point out that the flux density
ratio would also result in a reduction of the value of the cross
section for NFW lenses if we apply it in our calculations.
However, since we model the cluster-size lens halos as an NFW
profile, which contributes to lensing probabilities mainly at
larger image separations, where no confirmed lensing events are
found (i.e. the null results  of the JVAS/CLAS for
$6^{''}\leq\Delta\theta\leq 15^{''}$), if we adopt the mentioned
above usual form of an NFW lens cross section, the predicted
lensing probabilities will not be severely affected at smaller
image separations.

For galaxy-size SIS+SMBH modeled lenses, two images will always be
produced for all values of the source positions $|\vec{y}|$, and
larger $|\vec{y}|$ will produce a higher flux density ratio
$q_\mathrm{r}$. So a finite value of $q_\mathrm{r}$ will
definitely limit the source within a certain corresponding
position (which is also denoted by $y_\mathrm{cr}$) and hence
reduce the value of the cross section. The flux density ratio
$q_{\mathrm{r}}$ for the two images is the ratio of the
corresponding absolute values of magnifications (Schneider et
al.~\cite{schne}, Wu~\cite{wu}), $q_{\mathrm{r}}=|\mu_+/\mu_-|$,
where
\begin{equation}
\mu_+(y)=\left[\left(\frac{y}{x}\frac{dy}{dx}\right)_{x>0}\right]^{-1}
 \ \ \ \ \mathrm{and} \ \ \ \
\mu_-(y)=\left[\left(\frac{y}{x}\frac{dy}{dx}\right)_{x<0}\right]^{-1}.
\end{equation}
So $y_\mathrm{cr}$ is determined numerically by
$|\mu_+(y_{\mathrm{cr}})|=q_{\mathrm{r}}|\mu_-(y_{\mathrm{cr}})|$.

%
\begin{figure}
\resizebox{\hsize}{280pt}{\includegraphics{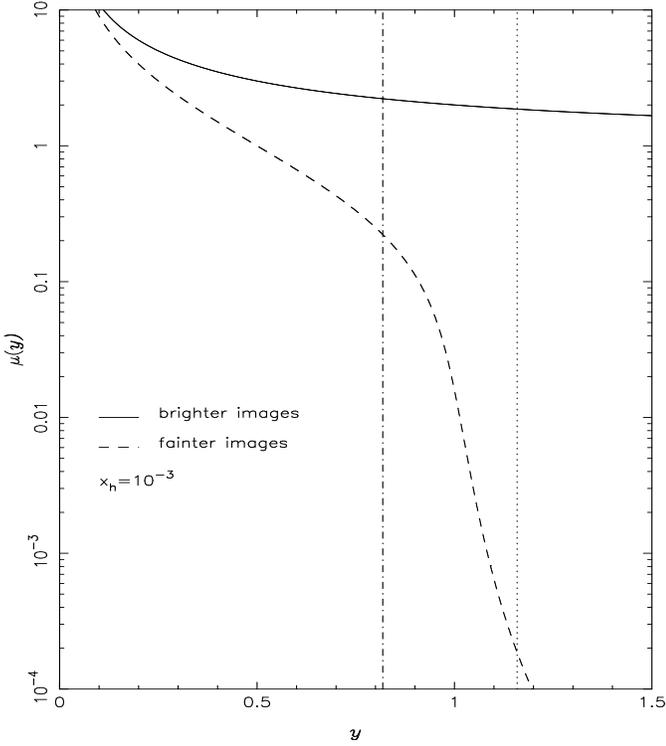}}
\caption{Image magnifications. Solid line: the brighter images
($\mu_+(y)$, $x>0$), dashed line: the fainter images ($\mu_-(y)$,
$x<0$). The left vertical line (dot-dashed) indicates the source
position $y_{\mathrm{cr}}$ at which $q_\mathrm{r}=10$, and the
right vertical line (dotted) indicates the position for
$q_\mathrm{r}=10^{4}$.} \label{fig2}
\end{figure}
%

We plot the magnifications (each as a function of source position
$y$) both for the brighter images and the fainter images in
Fig.~\ref{fig2}, with the source positions at which
$q_\mathrm{r}=10$ and $q_\mathrm{r}=10^4$ explicitly indicated,
and we have used a typical value $x_\mathrm{h}=10^{-3}$. The
sensitivity of the source position to the flux density ratio is
obvious. The cross section for images with a separation greater
than $\Delta\theta$ and a flux density ratio less than
$q_{\mathrm{r}}$ in the lens plane is (Schneider et
al.~\cite{schne}; Chen~\cite{chenb})
\begin{equation}
\begin{array}{ll}
\sigma(M,z)=& \pi\xi_0^2\vartheta(M-M_{\mathrm{min}}) \\
& \times \cases{ y_{\mathrm{cr}}^2,&for
$\Delta\theta\leq\Delta\theta_0$; \cr
y_{\mathrm{cr}}^2-y_{\Delta\theta}^2,&for
$\Delta\theta_0\leq\Delta\theta<\Delta\theta_{y_{\mathrm{cr}}}$;\cr
0,&for $\Delta\theta\geq\Delta\theta_{y_{\mathrm{cr}}}$,\cr}
\label{cross}
\end{array}
\end{equation}
where $\vartheta(x)$ is a step function, and $M_{\mathrm{min}}$ is
the minimum mass of halos above which lenses can produce images
with separations greater than $\Delta\theta$. From
Eq.(\ref{lenseq}), an image separation for any $y$ can be
expressed as $\Delta\theta(y)=\xi_0\Delta
x(y)/D_\mathrm{L}^\mathrm{A}$, where $\Delta x(y)$ is the image
separation in lens plane for a given $y$. So in Eq.(\ref{cross}),
the source position $y_{\Delta\theta}$, at which a lens produces
the image separation $\Delta\theta$, is the reverse of this
expression.  $\Delta\theta_0=\Delta\theta(0)$ is the separation of
the two images which are just on the Einstein ring;
$\Delta\theta_{y_{\mathrm{cr}}}=\Delta\theta(y_{\mathrm{cr}})$ is
the upper-limit of the separation above which the flux ratio of
the two images will be greater than $q_{\mathrm{r}}$. Note that
since $M_{\mathrm{DM}}$ ($M_{15}$) is related to $\Delta\theta$
through $\xi_0$ in equation (\ref{xi0}) and
$\sigma_v^2=GM_\mathrm{DM}/2r_\mathrm{vir}$, we can formally write
$M_{\mathrm{DM}}=M_{\mathrm{DM}}(\Delta\theta(y))$ and determine
$M_{\mathrm{min}}$ for galaxy-size lenses by
$M_{\mathrm{min}}=M_{\mathrm{DM}}(\Delta\theta(y_{\mathrm{cr}}))$.

\section{Discussion and conclusions}
\label{discuss}

\begin{figure}
\resizebox{\hsize}{280pt}{\includegraphics{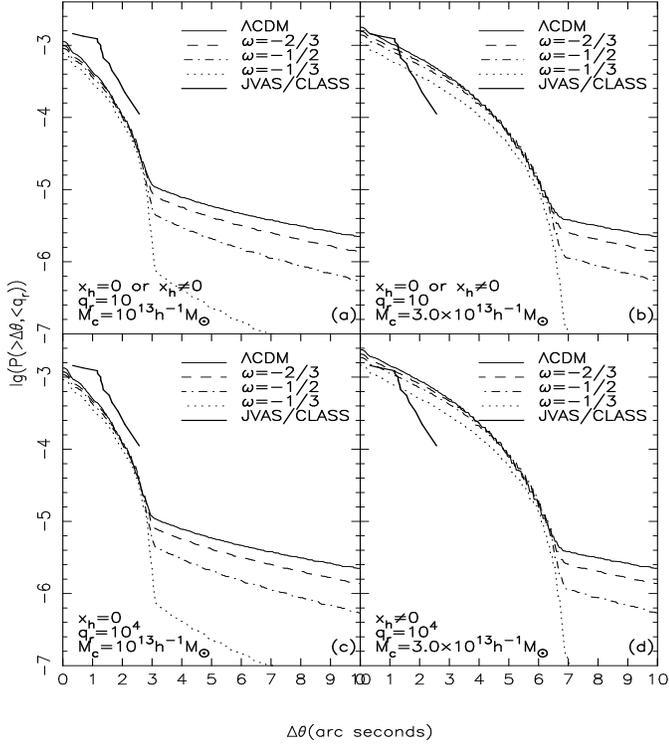}} \caption{The
integral lensing probabilities with image separations larger than
$\Delta\theta$ and flux density ratio less than $q_{\mathrm{r}}$,
for quasars at mean redshift $<z_\mathrm{s}>=1.27$ lensed by NFW
($M_\mathrm{DM}>M_\mathrm{c}$) and SIS+SMBH
($M_\mathrm{DM}<M_\mathrm{c}$) halos. The typically selected
values of the cosmic equation of state $\omega$ for each curve,
the flux density ratio $q_\mathrm{r}$ and cooling scale
$M_\mathrm{c}$ are explicitly indicated. The observed lensing
probabilities of JVAS/CLASS are plotted as a histogram in each
panel.} \label{fig3}
\end{figure}

The lensing probabilities predicted by Eq. (\ref{prob1}) and
calculated from the combined JVAS/CLASS data are compared in
Fig.~\ref{fig3}. Four cases with different flux density ratios
$q_\mathrm{r}$ or cooling scale $M_\mathrm{c}$ or with and without
central SMBH are investigated to show their effects on lensing
probabilities. In each case (see the corresponding panel of
Fig.~\ref{fig3}), four different values of the cosmic equation of
state $\omega$: $\omega=-1$ (cosmological constant),
$\omega=-2/3$, $\omega=-1/2$ and $\omega=-1/3$, are chosen to show
the constraints from lensing probabilities. As in our previous
work (Chen~\cite{chenb}), the observed lensing probabilities  are
calculated by
$P_{\mathrm{obs}}(>\Delta\theta)=N(>\Delta\theta)/8958$, where
$N(>\Delta\theta)$ is the number of lenses with separation greater
than $\Delta\theta$ in 13 lenses.

When flux density $q_\mathrm{r}=10$ (as allowed by the JVAS/CLASS
well-defined sample) and $M_\mathrm{c}=10^{13}h^{-1}M_{\sun}$, one
can see from the panel (a) of Fig.~\ref{fig3} that none of the
four values of $\omega$ is able to predict enough lensing
probabilities to match the observations. Note that the value of
the cooling mass scale used in this case is preferred by Kochanek
\& White (\cite{kocha}) and Li \& Ostriker (\cite{li}), and is
also close to the cutoff mass of halos below which cooling of the
corresponding baryonic component will lead to concentration of the
baryons to the inner parts of the mass profile (Rees \&
Ostriker~\cite{rees}; Blumenthal et al.~\cite{blume}; Porciani
\&Madau~\cite{porci}). So a larger value of $M_\mathrm{c}$
($>10^{13}h^{-1}M_{\sun}$) is needed for predictions to match
observations. When we attribute too few predicted probabilities in
panel (a) of Fig.~\ref{fig3} to the small value of $q_\mathrm{r}$
($=10$), the quite large value of $q_\mathrm{r}$ ($=10^4$) also
fails to match the observations, although the increment of the
probabilities is also obvious, as shown in panel (c). The reason
is that we have used different parameters and functions, e.g., the
sensitive concentration parameter $c_1$. Although it is not
explicitly displayed in the figure, our calculations indeed show
that if we set $q_\mathrm{r}=10^4$ and $M_\mathrm{c}=1.5\times
10^{13}h^{-1}M_{\sun}$, the predicted lensing probabilities for
$\omega=-1$ is able to match the observations quite well, which is
in agreement with the result obtained by Sarbu et al.
(\cite{sarbu}), since we have used almost the same parameters and
mass (and halo) functions. Note that when $q_\mathrm{r}$ is quite
large ($>10^3$), the contribution from the central galactic SMBH
cannot be ignored. In fact, it is just this contribution that
compensates the finite value effect of $q_\mathrm{r}=10^4$, which
make our results obtained in this case be the same as that of
Sarbu et al. (\cite{sarbu}), whose flux density value is infinite.

If we persist in  the flux density ratio allowed by the JVAS/CLASS
sample in our calculations, the cooling mass scale $M_\mathrm{c}$
must have the value of $3\times 10^{13}h^{-1}M_{\sun}$ for the
model with $\omega=-1$ to be able to match the observations, as
shown in panel (b). Such a value of $M_\mathrm{c}$ is close to
that used by Porciani \& Madau (\cite{porci}), however, they used
different parameters and, especially, they did not consider the
finite flux density ratio effect. Note that when $q_\mathrm{r}\leq
10$, the contributions from a galactic central SMBH can be
ignored, no matter what the value of $M_\mathrm{c}$ is, as shown
in the top two panels of Fig.~\ref{fig3}. This means that the
finite small flux density ratio will reduce the cross section of
SIS lenses considerably without a small central point mass. On the
other hand, our calculations shows that the central SMBH can be
detected only if $q_\mathrm{r}\geq 10^3$, this value is possible
for some confirmed radio-loud lens systems (Rusin \&
Ma~\cite{rusin2001}), but no sample suitable for analysis of the
lens statistics is available with such a high flux density ratio.
The sensitivity of lensing probability to the flux density ratio
is obvious when we compare panel (b) with panel (d) of
Fig.~\ref{fig3}. In the latter, $q_\mathrm{r}=10^4$, the effect of
which is close to that when flux density ratio is assumed to be
infinite (i.e., no constraints on the flux density ratio are taken
into account, as in most previous work; see Fig.~\ref{fig2}), and
other parameters, including $M_\mathrm{c}$, are the same as panel
(b). Obviously, for $q_\mathrm{r}=10^4$, the ``right" cosmic
equation of state to match observations is $\omega=-1/2$. The
present time deceleration parameter for a flat, dark energy
dominated Universe is $q_0=1/2+(3/2)\omega\Omega_\mathrm{Q}$, and
the accelerating Universe requires $q_0<0$, which immediately
gives $\omega<-1/2.1$ (we are using $\Omega_\mathrm{Q}=0.7$). So
the fit value of $\omega=-1/2$ in panel (d) is still within the
range required by the accelerating Universe. Clearly, when flux
density ratio $q_\mathrm{r}$ changes from $10$ to $10^4$, the
corresponding fit value of the cosmic equation of state  will
change from $\omega=-1$ to $\omega=-1/2$.

In summary, we revisit the two populations of lens halo model with
mass distribution NFW ($M_\mathrm{DM}>M_\mathrm{c}$) and SIS+SMBH
($M_\mathrm{DM}<M_\mathrm{c}$), to calculate lensing probabilities
in flat, low-density cosmological models with different cosmic
equations of state $\omega$. The finite flux density ratio effect
is significant. A low value of $q_\mathrm{r}$ can be compensated
for by raising the cooling mass scale $M_\mathrm{c}$ in fitting
the predicted lensing probabilities to JVAS/CLASS observations. To
determine the cosmic equation of state $\omega$, the uncertainty
in $M_\mathrm{c}$ must be resolved. The contributions from
galactic central SMBHs can be safely ignored.

\begin{acknowledgements}
I'm grateful to the anonymous referee for helpful suggestions.
This work was supported by the National Natural Science Foundation
of China under grant No.10233040.
\end{acknowledgements}
\appendix
\section{Lensing equation for SIS+SMBH}
In this Appendix, we give a detailed deduction of the lensing
equation for SIS+SMBH model. Choose the length scales in the lens
plane and the source plane to be
\begin{equation}
\xi_0=4\pi\left(\frac{\sigma_{v}}{c}\right)^2\frac{D_\mathrm{L}^\mathrm{A}
D_\mathrm{LS}^\mathrm{A}}{D_\mathrm{S}^\mathrm{A}} \ \ \
\mathrm{and} \ \ \
\eta_0=\xi_0\frac{D_\mathrm{S}^\mathrm{A}}{D_\mathrm{L}^\mathrm{A}},
\end{equation}
then the position vector in the lens plane and source plane is,
respectively, $\vec{\xi}=\vec{x}\xi_0$ and
$\vec{\eta}=\vec{y}\eta_0$. Note that we have used the different
length scales between SIS and NFW lenses. The dimensionless
surface mass density then is
\begin{equation}
\kappa(\vec{x})=\frac{M_{\bullet}\delta^2(\vec{x})}{\Sigma_\mathrm{cr}\xi^2_0}
+\frac{\sigma^2_{v}}{2G\Sigma_\mathrm{cr}\xi_0}\frac{1}{|\vec{x}|}.
\end{equation}
So the lensing equation is
\begin{equation}
y=x-m(x)/x,
\end{equation}
with
\begin{eqnarray}
m(x)&=&2\int^x_0\kappa(x')x'dx'  \nonumber \\
    &=&\frac{M_{\bullet}}{\pi\Sigma_\mathrm{cr}\xi^2_0}+x.
    \label{mx}
\end{eqnarray}
$y=|\vec{y}|$ and $x=|\vec{x}|$, and their values will be extended
to the negative for convenience in later calculations. We have
used the known formula $\delta^2(\vec{x})=\delta(x)/2\pi x$ to
derive the results in Eq. (\ref{mx}). In terms of  the radius of
the sphere of influence or Bondi accretion radius of a SMBH (e.g.,
Ferrarese \& Merritt~\cite{ferra}; Melia \& Falcke~\cite{melia};
Zhao et al.~\cite{zhao}) $r_\mathrm{h}=GM_\bullet/\sigma_v^2$, the
lensing equation can be rewritten as
\begin{equation}
y=x-\frac{x_\mathrm{h}}{x}-\frac{|x|}{x} \ \ \ \mathrm{with} \ \
\, x_\mathrm{h}=\frac{r_\mathrm{h}}{\pi\xi_0}. \label{lenseq}
\end{equation}
The mass of each galactic central SMBH $M_\bullet$ is related to
the mass of its host galaxy $M_\mathrm{DM}$ with the formula given
by Ferrarese (\cite{ferra2002})
\begin{equation}
\frac{M_{\bullet}}{10^{15}h^{-1}M_{\sun}}= 2.78\times
10^{-4}M_{15}^{1.57}, \label{bd}
\end{equation}
here we have defined a reduced  mass of the halos as
$M_{15}=M_\mathrm{DM}/(10^{15}\mathrm{h}^{-1}M_{\sun})$. On the
other hand, the mass of a SIS halo is defined the same way as for
NFW: $M_\mathrm{DM}=4\pi\int^{r_\mathrm{vir}}_0\rho(r)r^2dr$.
Since the mass density of a SIS halo is $\rho(r)=\sigma_v^2/2\pi
Gr^2$, the so-defined mass is related to the velocity dispersion
by $\sigma_v^2=GM_\mathrm{DM}/2r_\mathrm{vir}$. Generally, the
virial radius $r_\mathrm{vir}$ is cosmological model-dependent
according to its definition
$M_\mathrm{DM}=4\pi\delta_\mathrm{vir}\bar{\rho}(z)r_\mathrm{vir}^3/3$,
where $\delta_\mathrm{vir}$ is the well-known density contrast of
the virialized dark matter halos with the familiar value
$\delta_\mathrm{vir}=18\pi^2\approx 178$ for a standard cold dark
matter (SCDM) cosmology ($\Omega_\mathrm{m}=1$), and
$\bar{\rho}(z)$ is the mean mass density of the universe at
redshift $z$. We use $\delta_\mathrm{vir}=200$ in our actual
calculations,  since Jenkins et al. (\cite{jenki})  specifically
stated that their formula gives better fits to mass function
$f(M,z)$ with $\delta_\mathrm{vir}=178$ regardless of the
cosmological model (see Sarbu et al.~\cite{sarbu}). So the scaled
radius of the sphere of influence of the SMBH $x_\mathrm{h}$ can
be written explicitly
\begin{equation}
\begin{array}{cl}
x_\mathrm{h}=&5.18\times
10^{-4}\frac{M_{15}^{0.2367}}{D_\mathrm{R}^\mathrm{A}} \\
 &\times\left[\Omega_\mathrm{m}(1+z)^3+\Omega_\mathrm{Q}(1+z)^{3(1
+\omega)}\right]^{-2/3},
\end{array}
\end{equation}
where $D_\mathrm{R}^\mathrm{A}=
D_\mathrm{L}^\mathrm{A}D_\mathrm{LS}^\mathrm{A}
/D_\mathrm{S}^\mathrm{A}$ and we have used (for a flat Universe)
\begin{eqnarray}
\bar{\rho}(z)&=&\bar{\rho}(0)\frac{H^2(z)}{H^2_0} \nonumber \\
 &=&\bar{\rho}(0)\left[\Omega_\mathrm{m}(1+z)^3+\Omega_\mathrm{Q}
(1+z)^{3(1+\omega)}\right].
\end{eqnarray}


\begin{thebibliography}{99}
\bibitem[2003]{bahcall}
Bahcall, N. A., \& Bode, P. 2003, \apjl, 588, L1
\bibitem[1996]{barte}
Bartelmann, M. 1996, \aap, 313, 697
\bibitem[2002]{benson}
Benson, A. J., Lacey, C. G., Baugh, C. M., Cole S., \& Frnk, C. S.
2002, \mnras, 333, 156
\bibitem[2002]{bento} Bento, M. C., Bertolami, O., \& Sen, A. A.
2002, \prd, 66, 043507
\bibitem[2003]{bento03} Bento, M. C., Bertolami, O., \& Sen, A. A.
2003, \prd, 67, 063003
\bibitem[1986]{blume} Blumenthal, G. R., Faber, S. M., Flores, R.,
\& Primack, J. R. 1986, \apj, 301, 27
\bibitem[2000]{browne}
Browne, I.W., \& Meyers, S.T. 2000, in New Cosmological Data and
the Values of the Fundamental Parameters, IAU Symposium, 201, 47
\bibitem[2002]{browne02}
Browne, I. W. A., et al. 2002, preprint (astro-ph/0211069)
\bibitem[2001]{bullo} Bullock, J. S., Kolatt, T. S., Sigad, Y., et
al., 2001, \mnras, 321, 559
\bibitem[1998]{caldw} Caldwell, R. R, Dave, R., \& Steinhardt P. J.
1998, \prl, 80, 1582
\bibitem[1992]{carro}
Carroll, S. M., Press, W. H., \& Turner E. L. 1992, \araa, 30, 499
\bibitem[2003a]{chena} Chen, D. -M. 2003a, \aap, 397, 415
\bibitem[2003b]{chenb} Chen, D. -M. 2003b, \apj, 587, L55
\bibitem[2001]{cohn} Cohn, J. D., Kochanek, C. S., McLeod, B. A., \&
Keeton, C. R. 2001, \apj, 554, 1216
\bibitem[1999]{eisen} Eisenstein, D. J., \& Hu, W. 1999, \apj, 511, 5
\bibitem[2002]{ferra2002} Ferrarese, L. 2002, \apj, 578, 90
\bibitem[2000]{ferra}
Ferrarese, L., \& Merritt, D. 2000, \apjl, 539, L9
\bibitem[2000]{helbig}
Helbig, P. 2000, prepint, astro-ph/0008197
\bibitem[2001]{huter} Huterer, D., \& Turner, M. S. 2001, \prd, 64,
123527
\bibitem[2000]{jing}
Jing, Y. 2000, \apj, 535, 30
\bibitem[2001]{jenki}
Jenkins, A., et al. 2001, \mnras, 321, 372
\bibitem[1993]{kaufm} Kauffman, G., White, S. D. M., \& Guiderdoni, B.
1993, \mnras, 264, 201
\bibitem[2001]{keeton2001a}
Keeton, C. R. 2001, \apj, 561, 46
\bibitem[2002]{keeton02} Keeton, C. R. 2002, \apj, 582, 17
\bibitem[2001]{keeton2001b} Keeton, C. R., \& Madau, P. 2001, \apj,
549, L25
\bibitem[2001]{kocha}
Kochanek, C. S., \& White, M. 2001, ApJ, 559, 531
\bibitem[2002]{li}
Li, L. -X., \& Ostriker, J.P. 2002, ApJ, 566, 652
\bibitem[2003]{li2003}
Li, L. -X., \& Ostriker, J.P. 2003, ApJ, 595, 603
\bibitem[2003]{ma2003}
Ma, C. -P. 2003, ApJ, 584, L1
\bibitem[1999]{ma1999} Ma, C. -P., Caldwell, R. R.,  Bode, P., \&
Wang, L. 1999, \apjl, 52, L1
\bibitem[1998]{magor}
Magorrian, J., et al. 1998, \aj, 115, 2285
\bibitem[2000]{marlow} Marlow, D. R., Rusin, D., Jackson, N.,
Wilkinson, P. N., \& Browne, I. W. A. 2000, \aj, 119, 2629
\bibitem[1998]{mao} Mao, S., \& Schneider, P. 1998, \mnras, 295,
587
\bibitem[2001]{melia} Melia, F., \& Falcke, H. 2001, \araa, 39, 309
\bibitem[2003a]{meneg2003a} Meneghetti, M., Bartelmann, M., \&
Moscardini, L. 2003a, \mnras, 340, 105
\bibitem[2003b]{meneg2003b} Meneghetti, M., Bartelmann, M., \&
Moscardini, L. 2003b, \mnras, 346, 67
\bibitem[2001a]{merria}
Merritt, D., \& Ferrarese, L. 2001a, \apj, 547, 140
\bibitem[2001b]{merrib}
Merritt, D., \& Ferrarese, L. 2001b, \mnras, 320, L30
\bibitem[2001c]{merric}
Merritt, D., \& Ferrarese, L. 2001c, In: J. H. Knapen, J. E.
Beckman, I. Shlosman, T. J. Mahoney, eds., the Central Kpc of
Starbursts and AGN, ASP Conference Series, 249, 522
\bibitem[2002]{myers}
Myers, S. T., et al. 2002, \mnras, 334, 1
\bibitem[2002]{oguri}
Oguri, M. 2002, \apj, 580, 2
\bibitem[2003]{oguri2003a}
Oguri, M. 2003, \mnras, 339, L23
\bibitem[2003]{oguri2003b}
Oguri, M., Lee, J., \& Suto, Y. 2003, \apj, 599, 7
\bibitem[2000]{porci}
Porciani, C., \& Madau, P. 2000, ApJ, 532, 679
\bibitem[1974]{press74}
Press, W. H., \& Schechter, P. 1974, \apj, 187, 425
\bibitem[2001]{ravin}
Ravindranath, S., Ho, L., \& Filippenko, A. V. 2002, \apj, 566,
801
\bibitem[1977]{rees} Rees, M. J., \& Ostriker, J. P. 1977, \mnras, 179,
541
\bibitem[1997]{rix} Rix, H. -W., de Zeeuw, P. T., Cretton, N., van
der Marel, R. P., \& Carollo, C. 1997, \apj, 488, 702
\bibitem[1999]{roman} Romanowsky, A. J., \& Kochanek, C. S. 1999, \apj, 516,
18
\bibitem[2002]{rusin2002} Rusin, D. 2002, \apj, 572, 705
\bibitem[2001]{rusin2001} Rusin, D., \& Ma, C. -P. 2001, \apj, 549,
L33
\bibitem[2001]{sarbu}
Sarbu, N., Rusin, D., \& Ma, C.-P. 2001, \apj, 561, L147
\bibitem[2002]{sarzi}
Sarzi, M., Rix, H. -W., Shields, J. C., et al. 2002, \apj, 567,
237
\bibitem[1992]{schne}
Schneider, P., Ehlers, J., \& Falco, E. E. 1992, Gravitational
Lenses, Berlin: Springer-Verlag
\bibitem[1999]{somer} Somerville, R. S., \& Primack, J. R. 1999,
\mnras, 310, 1087
\bibitem[2002]{treu} Treu, T., \& Koopmans, L. V. E. 2002, \apj, 575,
87
\bibitem[2002]{wandel}
Wandel, A. 2002, \apj, 565, 762
\bibitem[1996]{wu}
Wu, X. -P. 1996, Fundamentals of Cosmic Physics, 17, 1
\bibitem[2002]{zhao} Zhao, H. S., Haehnelt, M. G., \& Rees, M. J. 2002,
New Astronomy, 7, 385
\end{thebibliography}
\end{document}